\documentclass[]{article}
\usepackage{lmodern}
\usepackage{amssymb,amsmath}
\usepackage{ifxetex,ifluatex}
\usepackage{fixltx2e} 
\ifnum 0\ifxetex 1\fi\ifluatex 1\fi=0 
  \usepackage[T1]{fontenc}
  \usepackage[utf8]{inputenc}
\else 
  \ifxetex
    \usepackage{mathspec}
  \else
    \usepackage{fontspec}
  \fi
  \defaultfontfeatures{Ligatures=TeX,Scale=MatchLowercase}
\fi
\IfFileExists{upquote.sty}{\usepackage{upquote}}{}
\IfFileExists{microtype.sty}{%
\usepackage{microtype}
\UseMicrotypeSet[protrusion]{basicmath} 
}{}
\usepackage[margin=1in]{geometry}
\usepackage{hyperref}
\hypersetup{unicode=true,
            pdftitle={dna2vec: Consistent vector representations of variable-length k-mers},
            pdfborder={0 0 0},
            breaklinks=true}
\usepackage{natbib}
\usepackage{longtable,booktabs}
\usepackage{graphicx,grffile}
\makeatletter
\def\maxwidth{\ifdim\Gin@nat@width>\linewidth\linewidth\else\Gin@nat@width\fi}
\def\maxheight{\ifdim\Gin@nat@height>\textheight\textheight\else\Gin@nat@height\fi}
\makeatother
\setkeys{Gin}{width=\maxwidth,height=\maxheight,keepaspectratio}
\IfFileExists{parskip.sty}{%
\usepackage{parskip}
}{
\setlength{\parindent}{0pt}
\setlength{\parskip}{6pt plus 2pt minus 1pt}
}
\providecommand{\tightlist}{%
  \setlength{\itemsep}{0pt}\setlength{\parskip}{0pt}}
\ifx\paragraph\undefined\else
\let\oldparagraph\paragraph
\renewcommand{\paragraph}[1]{\oldparagraph{#1}\mbox{}}
\fi
\ifx\subparagraph\undefined\else
\let\oldsubparagraph\subparagraph
\renewcommand{\subparagraph}[1]{\oldsubparagraph{#1}\mbox{}}
\fi

\let\rmarkdownfootnote\footnote%
\def\footnote{\protect\rmarkdownfootnote}

\usepackage{titling}

\newcommand{\subtitle}[1]{
  \posttitle{
    \begin{center}\large#1\end{center}
    }
}

  \title{dna2vec: Consistent vector representations of variable-length k-mers}
  \pretitle{\vspace{\droptitle}\centering\huge}
  \posttitle{\par}
\subtitle{Patrick Ng\\
\texttt{ppn3@cs.cornell.edu}}
  \author{}
  \preauthor{}\postauthor{}
  \date{}
  \predate{}\postdate{}

\usepackage{booktabs}  

\DeclareMathOperator*{\argmax}{arg\,max}

\begin{document}
\maketitle
\begin{abstract}
One of the ubiquitous representation of long DNA sequence is dividing it
into shorter k-mer components. Unfortunately, the straightforward vector
encoding of k-mer as a one-hot vector is vulnerable to the curse of
dimensionality. Worse yet, the distance between any pair of one-hot
vectors is equidistant. This is particularly problematic when applying
the latest machine learning algorithms to solve problems in biological
sequence analysis. In this paper, we propose a novel method to train
distributed representations of variable-length k-mers. Our method is
based on the popular word embedding model \emph{word2vec}, which is
trained on a shallow two-layer neural network. Our experiments provide
evidence that the summing of dna2vec vectors is akin to nucleotides
concatenation. We also demonstrate that there is correlation between
Needleman-Wunsch similarity score and cosine similarity of dna2vec
vectors.
\end{abstract}

\section{Introduction}\label{introduction}

The usage of k-mer representation has been a popular approach in
analyzing long sequence of DNA fragments. The k-mer representation is
simple to understand and compute. Unfortunately, its straightforward
vector encoding as a one-hot vector (i.e.~bit vector that consists of
all zeros except for a single dimension) is vulnerable to curse of
dimensionality. Specifically, its one-hot vector has dimension
exponential to the length of \(k\). For example, an 8-mer needs a bit
vector of dimension \(4^8 = 65536\). This is problematic when applying
the latest machine learning algorithms to solve problems in biological
sequence analysis, due to the fact that most of these tools prefer
lower-dimensional continuous vectors as input
\citep{suykens1999least, angermueller2016deep, turian2010word}. Worse
yet, the distance between any arbitrary pair of one-hot vectors is
equidistant, even though \texttt{ATGGC} should be closer to
\texttt{ATGGG} than \texttt{CACGA}.

\subsection{Word embeddings}\label{word-embeddings}

The Natural Language Processing (NLP) research community has a long
tradition of using bag-of-words with one-hot vector, where its dimension
is equal to the vocabulary size. Recently, there has been an explosion
of using \emph{word embeddings} as inputs to machine learning
algorithms, especially in the deep learning community
\citep{mikolov2013distributed, lecun2015deep, bengio2013representation}.
Word embeddings are vectors of real numbers that are distributed
representations of words.

A popular training technique for word embeddings, word2vec
\citep{mikolov2013efficient}, consists of using a 2-layer neural network
that is trained on the current word and its surrounding context words
(see Section \ref{sec-two-layer-neural-network}). This reconstruction of
context of words is loosely inspired by the linguistic concept of
\emph{distributional hypothesis}, which states that words that appear in
the same context have similar meaning \citep{harris1954distributional}.
Deep learning algorithms applied with word embeddings have had dramatic
improvements in the areas of machine translation
\citep{sutskever2014sequence, bahdanau2014neural, cho2014learning},
summarization \citep{chopra2016abstractive}, sentiment analysis
\citep{kim2014convolutional, dos2014deep} and image captioning
\citep{vinyals2015show}.

One of the most fascinating properties of word2vec is that its vector
arithmetic can solve semantic and linguistic analogies
\citep{mikolov2013linguistic, mikolov2013efficient}. They showed that
\(vec(king) - vec(man) + vec(woman) \approx vec(queen)\). In particular,
the analogy task \texttt{man:king\ ::\ woman:???} is interpreted as
finding a word \(w\) such that \(vec(king) - vec(man) + vec(woman)\) is
closest to \(vec(w)\) under cosine distance. Furthermore,
\citep{levy2014linguistic} showed that analogy works for past-tense
relation \(vec(capture) - vec(captured) \approx vec(go) - vec(went)\),
language-spoken-in relation
\(vec(france) - vec(french) \approx vec(mexico) - vec(spanish)\), as
well as geographical location
\(vec(Berlin) - vec(Germany) \approx vec(Paris) - vec(France)\).

\subsection{dna2vec: k-mer embeddings}\label{dna2vec-k-mer-embeddings}

In this paper, we present a novel method to compute distributed
representations of variable-length k-mers. These k-mers are consistent
across different lengths, i.e.~they lie in the same embedding vector
space. We embed k-mers of length \(3 \le k \le 8\), which is a space
consists of one k-mer per dimension (\(d = \sum_{k=3}^8 4^k\)), into a
continuous vector space of 100 dimensions.

The training method of our shallow two-layer neural network for dna2vec
is based on word2vec. BioVec \citep{asgari2015continuous} and seq2vec
\citep{kimothi2016distributed} have also applied the word2vec technique
to biological sequences. Although both techniques used a two-layer
neural network to train their embedding, our technique is a
generalization for variable-length \(k\).

\citep{needleman1970general} presented a method, now commonly known as
Needleman-Wunsch algorithm, for computing similarity of k-mers using a
dynamic programming scoring of global alignments. But the dynamic
programming nature of the algorithm makes the algorithm slow, with
quadratic time complexity to the length of the sequence. In Section
\ref{sec-nw-relationship}, we show that its cosine distance, in other
words angular distance, is related to Needleman-Wunsch distance of their
corresponding k-mers. In Section \ref{sec-analogy}, we provide evidence
that nucleotide concatenation analogy can be constructed with dna2vec
arithmetic.

The main contribution of this work includes:

\begin{itemize}
\tightlist
\item
  variable-length k-mer embedding model
\item
  experimental evidence that shows arithmetic of dna2vec vectors is akin
  to nucleotides concatenation
\item
  relationship between Needleman-Wunsch alignment and cosine similarity
  of dna2vec vectors
\item
  nucleotide concatenation analogy can be constructed with dna2vec
  arithmetic.
\end{itemize}

\section{Training dna2vec model}\label{training-dna2vec-model}

The training of dna2vec consists of four stages:

\begin{enumerate}
\def\labelenumi{\arabic{enumi})}
\tightlist
\item
  separate genome into long non-overlapping DNA fragments
\item
  convert long DNA fragments into overlapping variable-length k-mers
\item
  unsupervised training of an aggregate embedding model using a
  two-layer neural network
\item
  decompose aggregated model by k-mer lengths.
\end{enumerate}

\subsection{Stage 1: Long non-overlapping DNA
fragments}\label{stage-1-long-non-overlapping-dna-fragments}

We fragment the genome sequence based on gap characters (e.g.
\texttt{X}, \texttt{-}, etc). For our experiments using hg38 dataset,
the fragments were typically a couple of thousand nucleotides. To
introduce more entropy, we randomly choose to use the fragment's
reverse-complement.

\subsection{Stage 2: Overlapping variable-length
k-mers}\label{sec-overlapping-variable-length}

Given a DNA sequence \(S\), we convert the sequence \(S\) into
overlapping fixed length k-mer by sliding a window of length \(k\)
across \(S\). For example, we convert \texttt{TAGACTGTC} into five
5-mers: \texttt{\{TAGAC,\ AGACT,\ GACTG,\ ACTGT,\ CTGTC\}}. In the
variable-length case, we sample \(k\) from the discrete uniform
distribution \(\mathit{Uniform}(k_{low}, k_{high})\) to determine the
size of each window. For example, a sample of k-mers of
\(k \in \{3,4,5\}\) could be
\texttt{\{TAGA,\ AGA,\ GACT,\ ACT,\ CTGTC\}}.

Formally, given a sequence of length \(n\), \(S = (S_1, S_2, ..., S_n)\)
where \(S_i \in \{A,C,G,T\}\), we convert \(S\) into
\(\tilde{n} = n - k_{high} + 1\) number of k-mers:

\[f(S) = (S_{1:k_1}, S_{2:2 + k_2}, ... S_{\tilde{n}:\tilde{n} + k_{\tilde{n}}})\]
\[k_i \sim \mathit{Uniform}(k_{low}, k_{high})\]

where \(S_{a:b}\) is a shorthand for \((S_a, ..., S_b)\).

\subsection{Stage 3: Two-layer neural
network}\label{sec-two-layer-neural-network}

We use a shallow two-layer neural to train an aggregate DNA k-mer
embedding. The method is based on word2vec \citep{mikolov2013efficient}.
The word2vec algorithm has the options of continuous bag-of-words (CBOW)
or skip-gram. CBOW predicts the targeted word given the context, while
skip-gram predicts the context given the targeted word. The word2vec
homepage\footnote{\url{https://code.google.com/archive/p/word2vec/}}
claims that skip-gram is slower to train than CBOW, but skip-gram is
better for infrequent words. We use skip-gram for all our experiments.

Our dna2vec algorithm is trained by predicting the ``context''
surrounding a given targeted k-mer. The ``context'' is the set of
adjacent k-mers surrounding the targeted k-mer. For example, the context
of k-mer \texttt{GACT} would be \texttt{\{TAGA,\ AGA,\ ACT,\ CTGTC\}} in
our previous example from Section \ref{sec-overlapping-variable-length}.
For our experiments in this paper, we used a context size of 10 before
and after the targeted word, which amounts to predicting a total of 20
k-mers.

During training, either negative sampling or hierarchy softmax is
typically used to optimize the update procedure over all words. We used
negative sampling for all our experiments.

\subsection{Stage 4: Decompose aggregated model by k-mer
lengths}\label{stage-4-decompose-aggregated-model-by-k-mer-lengths}

We decompose the aggregate model by k-mer length to form
\(k_{high} - k_{low} + 1\) models. This decomposition is useful for the
searching of nearest neighbors, as we will discuss in Section
\ref{sec-nearest-neighbors}.

\section{Experiments}\label{experiments}

Our dna2vec\footnote{pre-trained dna2vec vectors available at
  \url{https://pnpnpn.github.io/dna2vec/} upon publication} was trained
with hg38 human assembly chr1 to chr22 \citep{rosenbloom2015ucsc}.
Specifically, they were downloaded from
\url{http://hgdownload.cse.ucsc.edu/downloads.html\#human}. We excluded
X and Y chromosomes, as well as mitochondrial and unlocalized sequences.

\subsection{Similarity and nearest
neighbors}\label{sec-nearest-neighbors}

For each vector arithmetic solution, we often compute its
\emph{n}-nearest k-mer neighbors. We define \emph{similarity} between
two dna2vec vectors \(v, w \in \mathbb{R}^d\) as the cosine similarity:

\[ sim(v, w) = \frac{v \cdot w} { \|v\| \|w\| } \]

The \emph{nearest-neighbor} of dna2vec vector \(v \in \mathbb{R}^d\) is
a k-mer computed with:

\begin{equation}
    \mathit{NearestNeighbor}_k(v) = \argmax_{s \in \{A,C,G,T\}^k} sim(v, vec(s))
    \label{eq:nearest-neighbor}
\end{equation}

Generally, the \(\mathit{nNearestNeighbors}_k(v)\) are the \(n\)-nearest
neighboring k-mers to vector \(v\).

\subsection{dna2vec arithmetic and nucleotide
concatenation}\label{dna2vec-arithmetic-and-nucleotide-concatenation}

We found that summing dna2vec embeddings is related to concatenating
k-mers. In Table \ref{tab:concat-basic}, we investigated this hypothesis
by adding dna2vec embeddings of two arbitrary k-mers and examining
whether their vector sum's neighbors overlap with their string
concatenation. The \emph{1-NN} column results were tallied using
Equation \eqref{eq:nearest-neighbor} and the other columns used
\emph{nNearestNeighbors} from Section \ref{sec-nearest-neighbors}. In
this experiment, string concatenation can come from both 5' and 3' ends.
For example, the following condition would be marked as a \emph{success}
for \emph{1-NN}:

\[ \mathit{NearestNeighbor}_6(vec(\mathtt{AAC}) + vec(\mathtt{TCT})) \in \{ \mathtt{AACTCT, TCTAAC} \}\]

Likewise, the following would be a \emph{success} for \emph{n-NN}:

\[ \mathit{nNearestNeighbors}_6(vec(\mathtt{AAC}) + vec(\mathtt{TCT})) \cap \{ \mathtt{AACTCT, TCTAAC} \} \ne \emptyset\]

\begin{longtable}[]{@{}llrrr@{}}
\caption{\label{tab:concat-basic} K-mers concatenation and dna2vec addition.
We took 1000 samples for each operand. For example, the first row is
aggregated from summing the dna2vec vectors of individual pairs of
arbitrary 3-mer and observing whether each of their string concatenation
overlaps with the vector sum's \emph{n}-nearest 6-mer
neighbors.}\tabularnewline
\toprule
\textbf{Operands} & \textbf{Concatenated} & \textbf{1-NN} &
\textbf{5-NN} & \textbf{10-NN}\tabularnewline
\midrule
\endfirsthead
\toprule
\textbf{Operands} & \textbf{Concatenated} & \textbf{1-NN} &
\textbf{5-NN} & \textbf{10-NN}\tabularnewline
\midrule
\endhead
3-mer + 3-mer & 6-mer & 28.7\% & 80.3\% & 94.6\%\tabularnewline
3-mer + 4-mer & 7-mer & 49.9\% & 90.4\% & 97.4\%\tabularnewline
3-mer + 5-mer & 8-mer & 53.9\% & 94.0\% & 98.4\%\tabularnewline
4-mer + 4-mer & 8-mer & 73.5\% & 96.8\% & 99.2\%\tabularnewline
\bottomrule
\end{longtable}

\subsection{Relationship to global alignment
similarity}\label{sec-nw-relationship}

All of the Needleman-Wunsch similarity score in this paper were computed
using Biopython's \texttt{align.globalxx} function, which used a match
score of 1, mismatch of 0 and gap penalty of 0.

\begin{figure}
\centering
\includegraphics{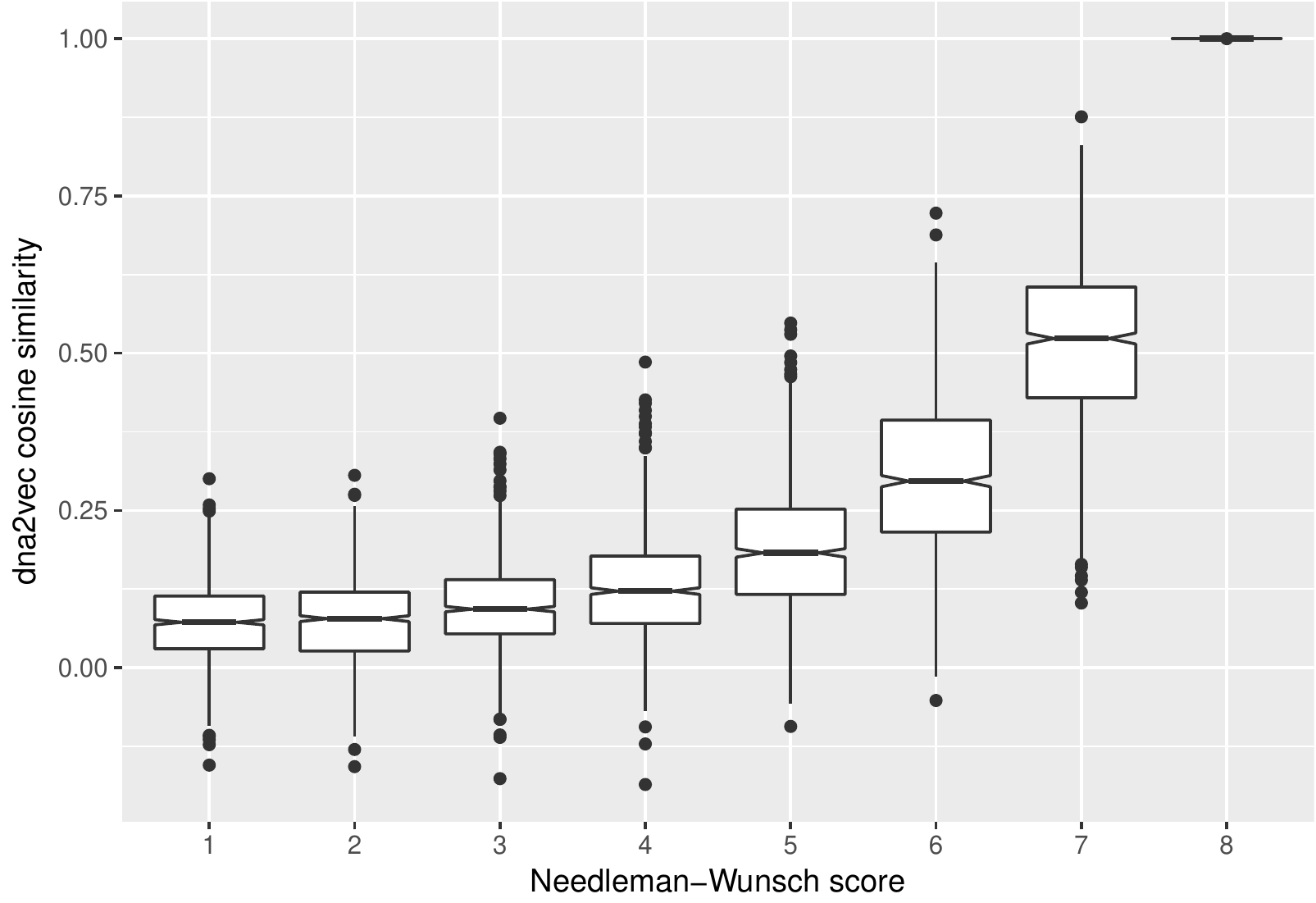}
\caption{\label{fig:nw-cosine-boxplot}Boxplot of Needleman-Wunsch score and dna2vec cosine
similarity. The lower and upper hinges are the 25 and 75 quartiles,
respectively. The Spearman's rank correlation coefficient is 0.831}
\end{figure}

In Figure \ref{fig:nw-cosine-boxplot}, we provided evidence that edit
distance between two arbitrary k-mers is correlated with the cosine
distance of their corresponding dna2vec vectors. We sampled 1000 pairs
of 8-mers for each Needleman-Wunsch score level and plot their
Needleman-Wunsch similarity score against dna2vec cosine similarity.

\begin{figure}
\centering
\includegraphics{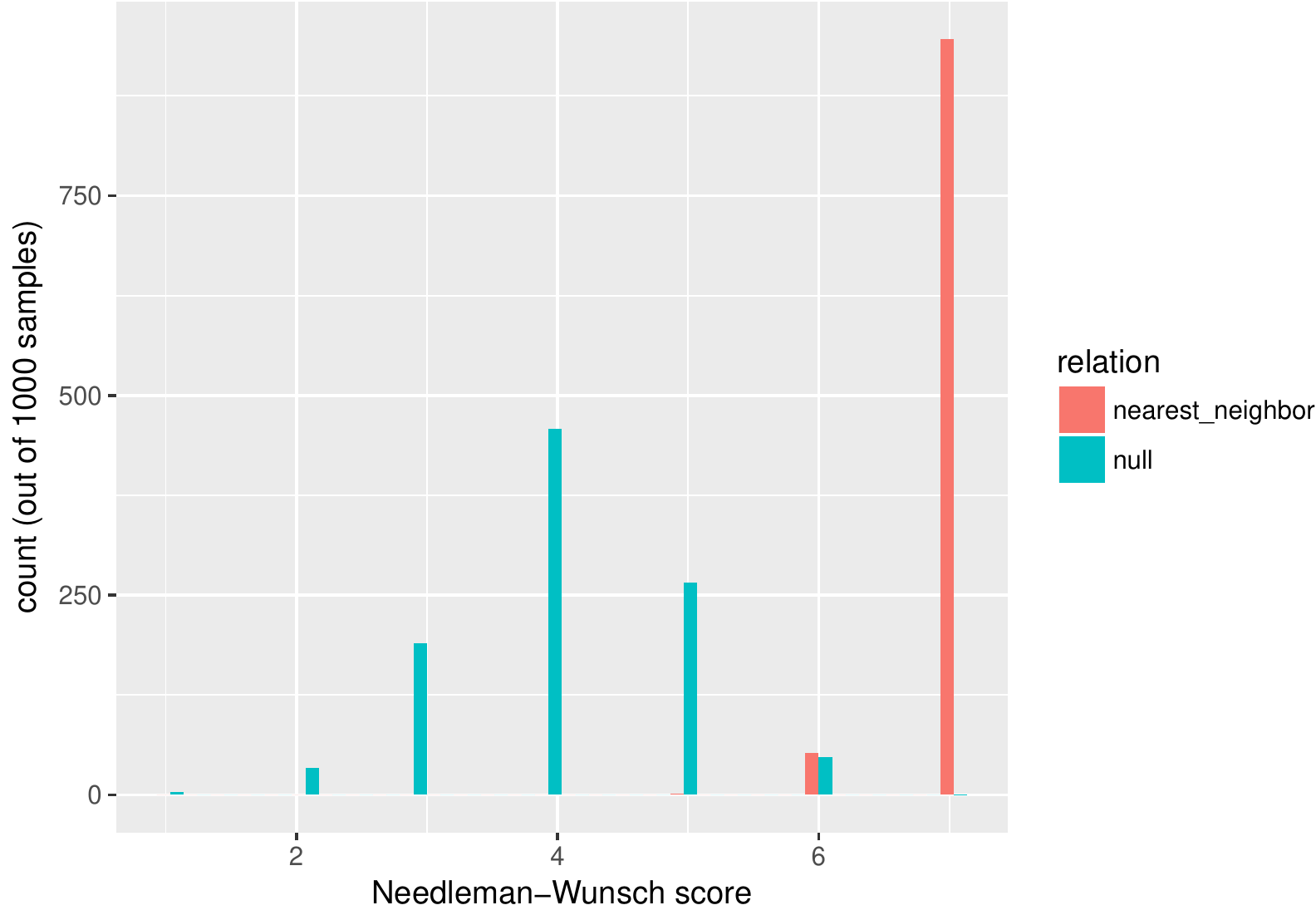}
\caption{\label{fig:nn-distribution}Global alignment score distribution of
nearest-neighbor. The \texttt{nearest-neighbor} distribution is
generated by computing Needleman-Wunsch score between 8-mer and its
nearest neighbor. The \texttt{null} distribution is from computing the
score between two random 8-mers.}
\end{figure}

In Figure \ref{fig:nn-distribution}, we compared the Needleman-Wunsch
similarity distribution of k-mer and its nearest dna2vec neighbor
against distribution of two random k-mers. Specifically, we sampled 1000
8-mers, found each of its nearest neighbor using Equation
\eqref{eq:nearest-neighbor}, and computed the Needleman-Wunsch score for
each pair. For the null distribution, we sampled 1000 pairs of random
8-mers. Thus we found evidence that the dna2vec nearest-neighbor
exhibits alignment similarity.

\subsection{Analogy of nucleotide concatenation}\label{sec-analogy}

We experimented with two types of nucleotide concatenation analogy:
\emph{strong} and \emph{weak} concatenations. Given two k-mers of the
same length, we define \emph{strong concatenation} as splicing
nucleotides on the same end (either 5' or 3' end) of the k-mers. An
example of 5-mer with 3-nucleotides snippet would be:

\[vec(\mathbf{AC}GAT) - vec(GAT) + vec(ATC) \approx vec(\mathbf{AC}ATC)\]

We relaxed the \emph{same end} restriction in the experiment \emph{weak
concatenation}, i.e.~the result can come from either end:

\[vec(\mathbf{AC}GAT) - vec(GAT) + vec(ATC) \in_{approx} \{ vec(\mathbf{AC}ATC), vec(ATC\mathbf{AC}) \}\]

Experimental samples were generated from randomly sampling two k-mers of
equal length and a nucleotide snippet (3 or 4 nucleotides) for
concatenation. For both strong and weak concatenation experiments, we
randomly selected either the 5' and 3' end to splice.

Table \ref{tab:insert-analogy} shows the summary of experimental results
from the two types of nucleotide concatenations. Particularly, we get
88\% accuracy with weak concatenation analogy of 8-mer and 4-nucleotides
snippet when considering 10-NN, as defined in Section
\ref{sec-nearest-neighbors}. Note that considering 30-nearest neighbors
is relatively small comparing to the space of all possible 6-mers, it is
merely 0.73\% of all possible 6-mers and 0.046\% of all possible 8-mers.

\begin{figure}
\centering
\includegraphics{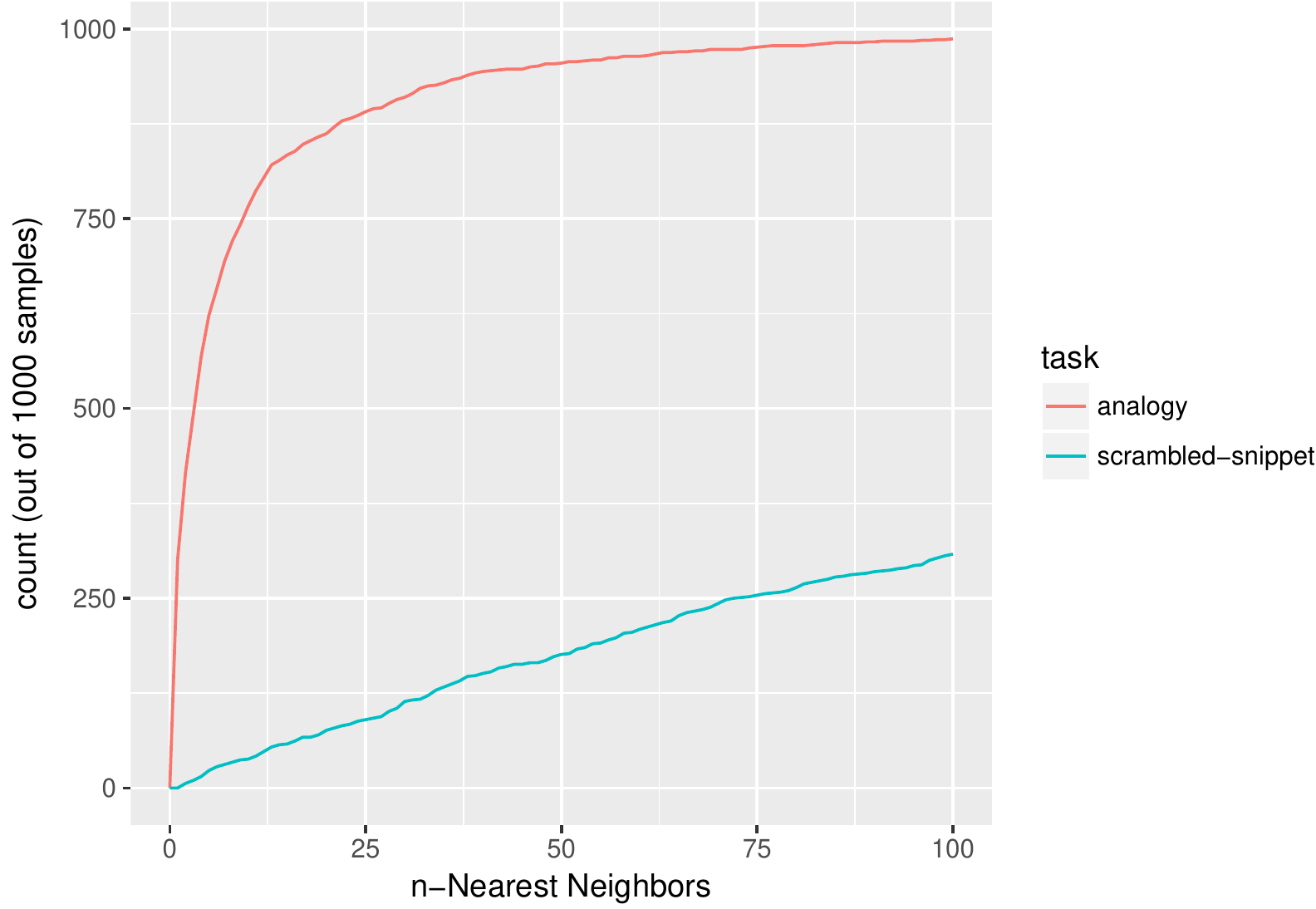}
\caption{\label{fig:cmf-analogy}Cumulative mass for analogy experiment of 8-mer
with 3-nt snippet. 1000 samples were generated with the
strong-concatenation analogy setup. We compared it with another 1000
samples using the \emph{scrambled-snippet} sampling procedure.}
\end{figure}

To confirm whether the arithmetic was actually extending a k-mer by the
snippet as oppose to similarity comparison, we compared the analogy
results with \emph{scrambled-snippet} experiments, which concatenated a
different random snippets in the answer case. As expected, the vector
arithmetic was significantly favoring the correct matching snippet
(\texttt{analogy} column) over a different random snippet
(\texttt{scrambled-snippet} column) in Figure \ref{fig:cmf-analogy} and
Table \ref{tab:insert-analogy}.

\begin{longtable}[]{@{}lllll@{}}
\caption{\label{tab:insert-analogy} Analogy Experiment. We analyzed two
types of analogies: \emph{weak} and \emph{strong} concatenation. 1000
samples were randomly generated for each type. For comparison, we
generated 1000 samples using scrambled-snippet sampling
strategy.}\tabularnewline
\toprule
\begin{minipage}[b]{0.19\columnwidth}\raggedright\strut
dimension\strut
\end{minipage} & \begin{minipage}[b]{0.19\columnwidth}\raggedright\strut
weak-concat

scrambled-snippet

5 / 10 / 30-NN\strut
\end{minipage} & \begin{minipage}[b]{0.19\columnwidth}\raggedright\strut
weak-concat

analogy

5 / 10 / 30-NN\strut
\end{minipage} & \begin{minipage}[b]{0.19\columnwidth}\raggedright\strut
strong-concat

scrambled-snippet

5 / 10 / 30-NN\strut
\end{minipage} & \begin{minipage}[b]{0.19\columnwidth}\raggedright\strut
strong-concat

analogy

5 / 10 / 30-NN\strut
\end{minipage}\tabularnewline
\midrule
\endfirsthead
\toprule
\begin{minipage}[b]{0.19\columnwidth}\raggedright\strut
dimension\strut
\end{minipage} & \begin{minipage}[b]{0.19\columnwidth}\raggedright\strut
weak-concat

scrambled-snippet

5 / 10 / 30-NN\strut
\end{minipage} & \begin{minipage}[b]{0.19\columnwidth}\raggedright\strut
weak-concat

analogy

5 / 10 / 30-NN\strut
\end{minipage} & \begin{minipage}[b]{0.19\columnwidth}\raggedright\strut
strong-concat

scrambled-snippet

5 / 10 / 30-NN\strut
\end{minipage} & \begin{minipage}[b]{0.19\columnwidth}\raggedright\strut
strong-concat

analogy

5 / 10 / 30-NN\strut
\end{minipage}\tabularnewline
\midrule
\endhead
\begin{minipage}[t]{0.14\columnwidth}\raggedright\strut
6-mer with 3-nt snippet\strut
\end{minipage} & \begin{minipage}[t]{0.17\columnwidth}\raggedright\strut
\emph{1.4 / 4 / 16\%}\strut
\end{minipage} & \begin{minipage}[t]{0.16\columnwidth}\raggedright\strut
47 / 69 / 95\%\strut
\end{minipage} & \begin{minipage}[t]{0.18\columnwidth}\raggedright\strut
\emph{0.6 / 1.8 / 9\%}\strut
\end{minipage} & \begin{minipage}[t]{0.21\columnwidth}\raggedright\strut
43 / 62 / 88\%\strut
\end{minipage}\tabularnewline
\begin{minipage}[t]{0.14\columnwidth}\raggedright\strut
7-mer with 3-nt snippet\strut
\end{minipage} & \begin{minipage}[t]{0.17\columnwidth}\raggedright\strut
\emph{2.4 / 6 / 16\%}\strut
\end{minipage} & \begin{minipage}[t]{0.16\columnwidth}\raggedright\strut
66 / 82 / 96\%\strut
\end{minipage} & \begin{minipage}[t]{0.18\columnwidth}\raggedright\strut
\emph{1.5 / 3.8 / 10\%}\strut
\end{minipage} & \begin{minipage}[t]{0.21\columnwidth}\raggedright\strut
61 / 76 / 92\%\strut
\end{minipage}\tabularnewline
\begin{minipage}[t]{0.14\columnwidth}\raggedright\strut
8-mer with 3-nt snippet\strut
\end{minipage} & \begin{minipage}[t]{0.17\columnwidth}\raggedright\strut
\emph{3 / 6 / 19\%}\strut
\end{minipage} & \begin{minipage}[t]{0.16\columnwidth}\raggedright\strut
67 / 82 / 95\%\strut
\end{minipage} & \begin{minipage}[t]{0.18\columnwidth}\raggedright\strut
\emph{2.3 / 3.8 / 11\%}\strut
\end{minipage} & \begin{minipage}[t]{0.21\columnwidth}\raggedright\strut
62 / 77 / 91\%\strut
\end{minipage}\tabularnewline
\begin{minipage}[t]{0.14\columnwidth}\raggedright\strut
8-mer with 4-nt snippet\strut
\end{minipage} & \begin{minipage}[t]{0.17\columnwidth}\raggedright\strut
\emph{0.7 / 1.4 / 3\%}\strut
\end{minipage} & \begin{minipage}[t]{0.16\columnwidth}\raggedright\strut
75 / 88 / 98\%\strut
\end{minipage} & \begin{minipage}[t]{0.18\columnwidth}\raggedright\strut
\emph{0.3 / 1.0 / 2.4\%}\strut
\end{minipage} & \begin{minipage}[t]{0.21\columnwidth}\raggedright\strut
69 / 83 / 95\%\strut
\end{minipage}\tabularnewline
\bottomrule
\end{longtable}

\subsection{Implementation Details}\label{implementation-details}

We will make our code and data available at
\url{https://pnpnpn.github.io/dna2vec/} upon publication. The two-layer
neural network training method described in Section
\ref{sec-two-layer-neural-network} was implemented using \texttt{gensim}
framework \citep{gensim}. We used gensim's \texttt{Word2vec} class with
parameters \texttt{sg=1} and \texttt{window=10}, which specified the
usage of skipgram model and the half-size of the context window as 10,
respectively. All of trained dna2vec vectors used in this paper has
dimension size of 100. Since the window sliding step in Section
\ref{sec-overlapping-variable-length} is stochastic in terms of
variable-length \emph{k}, we could essentially generate more training
data by looping through the complete genomic sequence data with multiple
passes, which we called \emph{epochs}. The dna2vec model used in this
paper was trained with 10 epochs. The training step took over 3 days
using gensim parameter \texttt{workers=4} on a 2.66 GHz Quad-Core Intel
Xeon with 8GB memory.

\section{Discussion}\label{discussion}

In this work, we presented a novel method for training distributed
representations of k-mers. We demonstrated that our dna2vec embeddings
can represent variable-length k-mers in a consistent fashion via
nucleotide concatenation experiments. We provided experimental evidence
showing that the arithmetic of dna2vec vectors is akin to nucleotides
concatenation. We also showed that Needleman-Wunsch similarity score
between two arbitrary k-mers is correlated with the cosine distance of
their corresponding dna2vec vectors. As for future work, due to the fact
that many machine learning algorithms require fixed-length continuous
vectors as input, we will explore the application of dna2vec with
machine learning techniques on biological sequence analysis.

\bibliography{book.bib}

\end{document}